%% file: EXO-11-020_temp.tex
\pdfoutput=1

\documentclass[11pt,twoside,a4paper,cmspaper,final,collab]{cms-tdr}
\def\svnVersion{175806}\def\cmsCernNo{2012-170}\def\cmsCernDate{\today}\def\cmsMessage{Submitted to the Journal of High Energy Physics}

\begin{document}\cmsNoteHeader{EXO-11-020}

\hyphenation{had-ron-i-za-tion}
\hyphenation{cal-or-i-me-ter}
\hyphenation{de-vices}

\newcommand{\livetime}{246}
\newcommand{\gluinoMassLimit}{640}
\newcommand{\stopMassLimit}{340}

\RCS$Revision: 133609 $
\RCS$HeadURL: svn+ssh://svn.cern.ch/reps/tdr2/papers/EXO-11-020/trunk/EXO-11-020.tex $
\RCS$Id: EXO-11-020.tex 133609 2012-06-29 10:58:04Z alverson $
\cmsNoteHeader{EXO-11-020} 
\title{Search for stopped long-lived particles produced in pp collisions at $\sqrt{s}=7$ TeV}

\date{\today}

\abstract{
A search has been performed for long-lived particles that have stopped in the CMS detector, during 7\TeV proton-proton operations of the CERN LHC.  The existence of such particles could be inferred from observation of their decays when there were no proton-proton collisions in the CMS detector, namely during gaps between LHC beam crossings. Using a data set in which CMS recorded an integrated luminosity of 4.0$\fbinv$, and a search interval corresponding to 246~hours of trigger live time, 12 events are observed, with a mean background prediction of $8.6 \pm 2.4$ events. Limits are presented at 95\% confidence level on long-lived gluino and stop production, over 13 orders of magnitude of particle lifetime.  Assuming the ``cloud model'' of R-hadron interactions, a gluino with mass below 640\GeV and a stop with mass below 340\GeV are excluded, for lifetimes between 10\mus and 1000\unit{s}.}

\hypersetup{%
pdfauthor={CMS Collaboration},%
pdftitle={Search for stopped long-lived particles produced in pp collisions at sqrt(s)=7 TeV},%
pdfsubject={CMS},%
pdfkeywords={CMS, physics, stopped gluino}}

\maketitle 

 \section{Introduction}
 \label{sec:intro}

Heavy long-lived particles are predicted by many extensions of the standard model, including supersymmetric models~\cite{Dimopoulos:1996vz, Baer:1998pg,Jittoh:2005pq}, ``hidden valley'' scenarios~\cite{Strassler:2006im}, grand-unified theories~\cite{Arvanitaki:2008hq} and split supersymmetry  models~\cite{ArkaniHamed:2004fb}. In models in which their production proceeds via the strong interaction, relatively large cross sections at the Large Hadron Collider (LHC) are predicted~\cite{Dawson:1983fw,Beenakker:1996ch,Plehn:2005cq,Beenakker:2009ha}.  While there are astrophysical constraints on the lifetime of such particles~\cite{Arvanitaki:2005fa}, these are imprecise owing to uncertainty about their interactions. In this article, we look for evidence of long-lived particles that stop in the Compact Muon Solenoid (CMS) detector and decay in the quiescent periods between LHC beam crossings.  Previous collider searches have used this method to set limits on the gluino ($\PSg$) lifetime and mass~\cite{Abazov:2007ht,Khachatryan:2010uf,Aad:2012zn,Farrar:2010ps}.  We now expand the search to include scalar top quarks, known as stops ($\widetilde{\rm t}$).

If long-lived gluinos or stops are produced at the LHC, they are expected to hadronize into $\PSg{\rm g}, \PSg{\rm q}\overline{\rm q}, \PSg{\rm qqq}$, $\widetilde{\rm t}\overline{\rm q}, \widetilde{\rm t}{\rm qq}$ states called ``R-hadrons''~\cite{Fayet:1977yc,Fayet:1978qc,Farrar:1978xj}.  These R-hadrons lose energy via nuclear interactions and, if charged, via ionisation, as they traverse the CMS detector.  The R-hadrons that are produced with sufficiently low velocity will come to rest in the detector volume~\cite{Arvanitaki:2005nq}, preferentially in the densest detector elements, the calorimeters.  These stopped particles may subsequently decay, which may result in an energy deposit in the CMS calorimeters similar to that produced by a jet.  If these decays occur out-of-time with respect to the proton-proton collisions producing the parent particles, the observation of such asynchronous events would constitute unambiguous evidence of new physics. This search is complementary to analyses~\cite{Khachatryan:2011ts,Aad:2011hz,Chatrchyan:2012sp} in which the long-lived particle signature is based on energy loss measurements in the charged-particle tracking system and timing information. Because this analysis relies only on calorimetry, it is sensitive to non-relativistic particles with velocity $\beta \le 0.3$, for which searches based on tracking information have negligible acceptance. The use of calorimetry also provides sensitivity to R-hadrons even in enhanced ``charged-flipping'' scenarios~\cite{Mackeprang:2009ad}, where the R-hadron becomes neutral and track reconstruction may be difficult or impossible. Finally, the stopped particle method would allow an experimental measurement of the particle lifetime and other properties~\cite{Graham:2011ah}.

Section~\ref{sec:det} provides an overview of the CMS experimental apparatus and the data sets used for the analysis. In Section~\ref{sec:sim} we describe the dedicated simulation procedure for the signal, and in Section~\ref{sec:sel}, the online and offline event selection.  The background estimation methods are described in Section~\ref{sec:backgrounds}.  The systematic uncertainties are listed in Section~\ref{sec:syst} and the results are given in Section~\ref{sec:results}.  Finally, in Section~\ref{sec:exclusion}, we interpret the results as an excluded region in the $m_{\PSg}$ - $m_{\widetilde{\chi}^{0}}$ plane.

\section{The CMS detector}
\label{sec:det}

The central feature of the CMS apparatus is a superconducting solenoid, of 6\unit{m} internal diameter, providing a magnetic field of 3.8\unit{T}. Within the field volume are a silicon pixel and strip tracker, a lead tungstate crystal electromagnetic calorimeter (ECAL), and a brass/scintillator hadron calorimeter (HCAL).  Hybrid photodiodes (HPDs) are used for light collection from the HCAL scintillators.  Muons are measured in gas-ionisation detectors embedded in the steel return yoke; drift tubes (DT) and resistive plate chambers (RPC) provide coverage in the barrel, while cathode strip chambers (CSC) and RPC provide coverage in the endcaps. Extensive forward calorimetry complements the coverage provided by the barrel and endcap detectors.  CMS uses a right-handed coordinate system, with the origin at the nominal interaction point, the $x$ axis pointing to the centre of the LHC, the $y$ axis pointing up (perpendicular to the LHC plane), and the $z$ axis along the anticlockwise-beam direction. The polar angle, $\theta$, is measured from the positive $z$ axis and the azimuthal angle, $\phi$, is measured in the $x$-$y$ plane.  The pseudorapidity, $\eta$, is defined as $\eta=-\ln[\tan(\theta/2)]$.  In the region $\abs{\eta}< 1.74$, the HCAL cells have widths of 0.087 in $\eta$ and 0.087\ in $\phi$. In the $\eta$-$\phi$ plane, and for $\abs{\eta}< 1.48$, the HCAL cells map onto $5 \times 5$ ECAL crystals arrays to form calorimeter towers projecting radially outwards from close to the nominal interaction point. At larger values of $\abs{\eta}$, the size of the towers increases and the matching ECAL arrays contain fewer crystals. Within each tower, the energy deposits in ECAL and HCAL cells are summed to define the calorimeter tower energies, subsequently used to provide the energies and directions of hadronic jets. For this search, jets are reconstructed offline from the energy deposits in the calorimeter towers, clustered by an iterative cone algorithm~\cite{Chekanov:2002rq} with a cone radius of 0.5 in $(\eta,\phi)$ space. Muons, which are used to reject cosmic ray backgrounds, are measured in the pseudorapidity range $\abs{\eta}< 2.4$.
A more detailed description of the CMS apparatus can be found in Ref.~\cite{CMS:2008zzk}.

The search is performed using data recorded in 2011 during regular periods of LHC colliding beams (fills).  Several fills taken towards the end of 2011 were not included in the search because the LHC vacuum was relatively low during these periods, which may have resulted in beam-gas backgrounds to which the search would have been sensitive.  During the period analysed, CMS recorded an integrated luminosity of 4.0\fbinv, although it should be noted that the integrated luminosity to which this search is sensitive depends on the particle lifetime.  Since we search for stopped particle decays in the gaps between colliding bunches, the fraction of time available for the search depends on the number (and spacing) of proton bunches in the LHC.  For the fills analysed, the number of bunches per beam varied from 228 to 1380.  During event selection, we veto any event within two 25\unit{ns} LHC clock cycles (BX) of a bunch passing through CMS.  In addition, physics triggers are inhibited for a short time at the end of the LHC revolution period (the ``LHC orbit''), in which calibration triggers are taken.  After accounting for these periods, between 85\%, for 228 bunch fills, and 16\%, for 1380 bunch fills, of the LHC orbit is available for the search.  The total time during which the analysis is sensitive to decays of stopped particles (the ``live time'') is \livetime~hours.  For measurement of the background due to instrumental noise, we use a control sample taken between March and August 2010, comprising 249~hours of total live time.  The integrated luminosity delivered in the control sample (3.6\pbinv) is sufficiently small that signal contamination can be neglected.

\section{Signal simulation}
\label{sec:sim}
\begin{figure}[th]
 \centering
   \includegraphics[width=0.8\textwidth]{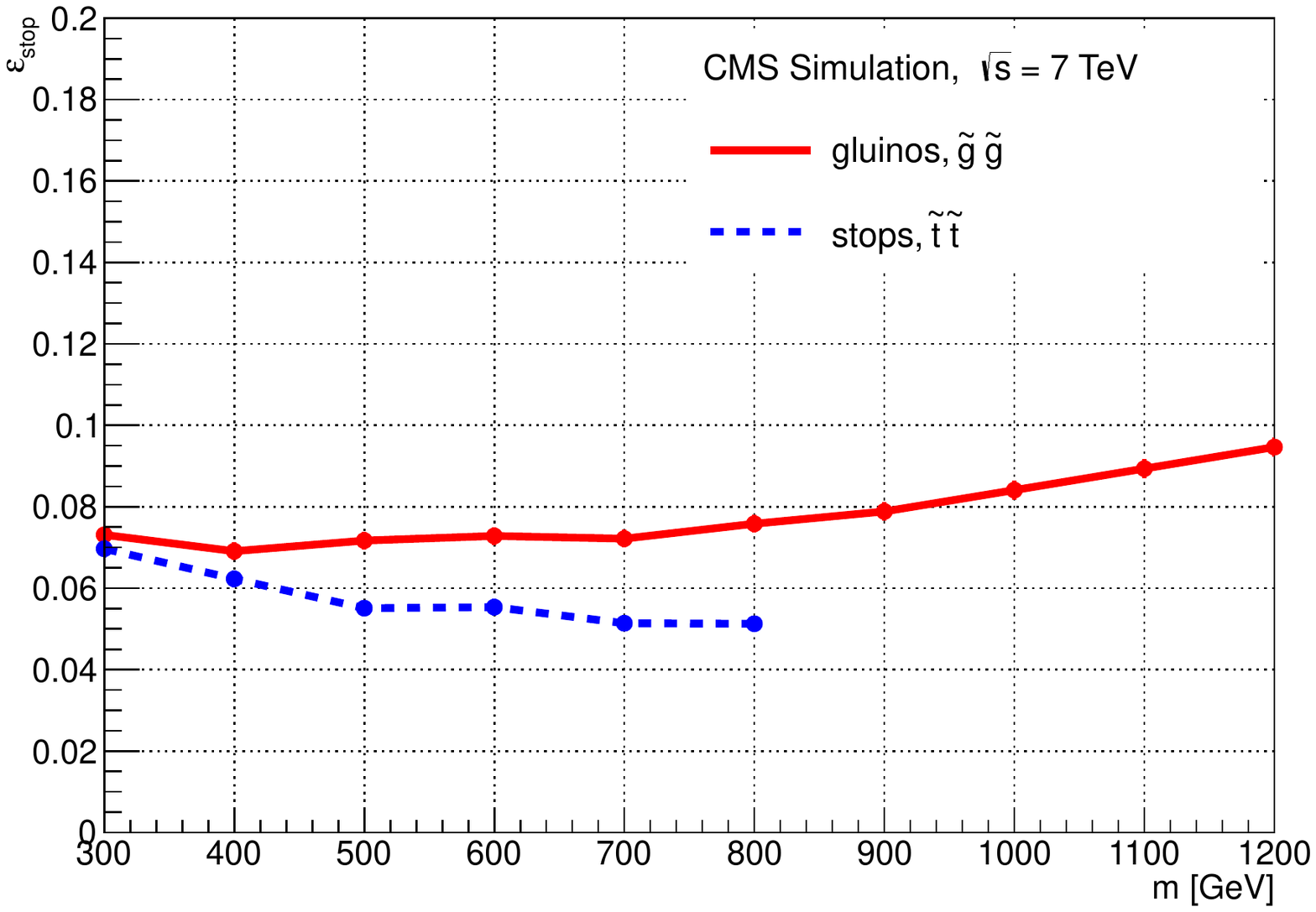}
   \caption{Probability for one of a pair of produced long-lived particles to stop in the barrel calorimeter (ECAL or HCAL), as a function of the particle mass.}
   \label{fig:stoppingProbability}
\end{figure}

The same factorized simulation is employed as in our previous publication ~\cite{Khachatryan:2010uf}.  In the first phase, ${\rm q}\overline{\rm q}\rightarrow\PSg\PSg$, ${\rm gg}\rightarrow\PSg\PSg$, ${\rm q}\overline{\rm q}\rightarrow\widetilde{\rm t}\widetilde{\rm t}$, and ${\rm gg}\rightarrow\widetilde{\rm t}\widetilde{\rm t}$ events are generated at $\sqrt{s} = 7$\TeV using \PYTHIA~\cite{Sjostrand:2006za} to simulate particle pair production, together with their subsequent hadronisation into R-hadrons~\cite{Fairbairn:2006gg}.  In this phase the gluino or stop is defined to be stable. We simulate gluino masses in the range $m_{\PSg} = 300$--1200\GeV, and stop masses in the range $m_{\widetilde{\rm t}} = 300$--800\GeV.   A modified \GEANTfour simulation~\cite{Agostinelli:2002hh} that implements a phase-space driven ``cloud model'' of strong R-hadron interactions with matter~\cite{Kraan:2004tz,Mackeprang:2006gx}, referred to as the ``generic'' model in Ref.~\cite{Aad:2012zn}, is used to simulate the interaction of these R-hadrons with the CMS detector and to record the final locations of those R-hadrons that stop in the detector.  As shown in Fig.~\ref{fig:stoppingProbability}, the probability of at least one R-hadron to stop in the CMS detector was determined to be in the range 0.05--0.07, for both gluinos and stops in the mass range studied.  

In the next phase, we simulate the decay of the stopped R-hadron.  Implicit in our factorized approach is the assumption that the decay time is much greater than the time required to stop the R-hadron.  We again use \PYTHIA, this time to produce a R-hadron at rest; we then translate from the nominal vertex position to the recorded stopping location and decay the constituent gluino or stop instantaneously.  We assume a 100\% branching fraction for $\PSg \rightarrow {\rm g}\,\widetilde{\chi}^0_1$, which dominates in split supersymmetry for the range of gluino mass to which the search is sensitive~\cite{Gambino:2005eh}, although it should be noted that our method is also sensitive to final states with additional jets such as $\PSg \rightarrow {\rm q \overline{q}}\widetilde{\chi}^0_1$.  Similarly, we assume a 100\% branching fraction for $\widetilde{\rm t} \rightarrow {\rm t}\,\widetilde{\chi}^0_1$. The kinematics of the R-hadron decay is dominated by the properties of the gluino or stop, and neutralino, as the spectator quarks do not play a significant role.  These spectator quarks cannot be ignored, however, as they participate in the hadronisation of the gluon or quarks produced in the decay.  We developed and implemented a custom decay table to correctly account for the colour flow in these decays.  Finally, we use a specialised Monte Carlo simulation that uses the measured delivered luminosity profile and the LHC filling scheme to simulate the time profile of stopping particle production.  By randomly generating decay times according to the exponential distribution expected for a given lifetime hypothesis, and using the record of good data-taking periods and the LHC filling scheme, the simulation program calculates the fraction of stopped particle decays that occur during a triggerable interval, and hence determines an effective luminosity for each lifetime hypothesis.

\section{Event selection}
\label{sec:sel}

In order to record decays of particles during gaps between the proton bunches that comprise the LHC beam, we use an updated version of the dedicated calorimeter trigger employed in our previous search~\cite{Khachatryan:2010uf}.  This trigger uses information from the two beam position and timing (BPTX) monitors that are positioned 175~m from the centre of CMS, along the beam axis on each side, and that produce a signal when an LHC proton bunch passes.  We require a jet trigger together with the condition that both BPTX signals are below threshold, ensuring that the trigger will not select particles from pp collisions or from beam-gas interactions of protons in unpaired bunches.  We also require that at most one BPTX produces a signal in a window $\pm1$~BX around the triggered event.  This reduces the trigger rate due to lower intensity, out of time, ``satellite'' bunches that accompany the colliding protons.  The energy deposition in the calorimeter of a jet from the R-hadron decay is sufficiently similar to those of jets originating directly from pp collisions that a calorimeter jet trigger can be used.  At the hardware trigger level (L1), the jet transverse energy, $E_{T}$, which is calculated assuming the jet was produced at the nominal interaction point, is required to be greater than 32\GeV, while in the software trigger (HLT) the jet energy is required to be greater than 50\GeV.  At both L1 and HLT the pseudorapidity of the jet, $|\eta_\text{jet}|$, is required to be less than 3.0.  Finally, the trigger vetoes any event that is accompanied by a L1 endcap muon beam-halo trigger, again within a $\pm1$~BX window.

The offline event selection, described below, is designed to reject backgrounds related to the passage of beam through CMS, as well as backgrounds from cosmic rays and from instrumental noise, while retaining good signal efficiency.  We describe in turn the selection criteria that reject each source of background.

Beam related backgrounds include out-of-time pp collisions, beam-gas interactions from unpaired proton bunches, and beam-halo.  In order to reject events related to an unpaired bunch passing through CMS, events in which either BPTX is over threshold are vetoed, in a $\pm2$~BX window around the trigger.  This criterion also rejects other beam backgrounds, such as beam halo, and mis-timed jet triggers from in-time pp collisions.  The most significant remaining beam-related background arises from beam halo.  Beam-halo muons that result in energy deposits in the HCAL are well-reconstructed as track segments in the muon endcap CSC system, so we reject any event containing a CSC segment.  Finally, to ensure that no out-of-time pp collision events due to satellite bunches contaminate the search sample, events with one or more reconstructed primary vertices are rejected.

A small fraction of cosmic rays traversing CMS deposit significant energy in the calorimeters.  To remove such background, we veto events which contain one or more reconstructed muons.  Because of inefficiencies in the track reconstruction of cosmic rays, we also veto events which contain more than two DT segments, or more than two RPC hits.  In the muon endcap, we require the hits in each RPC hit pair to be separated by $\Delta\phi<0.4$, and in the muon barrel by $\Delta z<40$\unit{cm}.

In addition to beam-related backgrounds and cosmic rays, instrumental noise can also mimic our signal.  To combat this background, standard calorimeter cleaning and noise rejection criteria~\cite{Chatrchyan:2009hy} are applied.  We restrict our search to jets in the less noisy central HCAL, requiring that the most energetic jet in the event has $ |\eta_{\text{jet}}| < 1$.  This requirement has been tightened from the value used in Ref.~\cite{Khachatryan:2010uf} in order to improve the rejection of beam-halo events, which tend to include calorimeter deposits at large pseudorapidity.  A jet with reconstructed energy above 70\GeV is required, above the trigger efficiency turn-on region.  To remove events due to noise in a single HCAL channel, events with more than 90\% of the energy deposited in three or fewer calorimeter towers are vetoed.  We also require that the leading jet has at least 60\% of its energy contained in fewer than 6 towers.  To suppress noise from HPD discharges~\cite{Chatrchyan:2009hy}, events with 5 or more of the leading towers at the same azimuthal angle, or with more than 95\% of the jet energy contained within towers at the same azimuthal angle, are rejected.  Finally, we compute the largest fraction of jet energy contained within a single strip of towers at constant $\phi$, and we require that this fraction is less than 95\%.

The HCAL electronics have a well-defined time response to charge deposits generated by showering particles. Analog signal pulses produced by these electronics are sampled at 40\unit{MHz}, synchronised with the LHC clock.  These pulses are read out over ten BX samples grouped about the pulse maximum. A pulse resulting from real energy deposition in the calorimeters (``a physical pulse'') has some notable properties, which we use to distinguish it from noise pulses.  Physical pulses have a clear peak containing a large fraction of the energy ($E_{\text{peak}}$), significant energy in one bunch crossing before the peak ($E_{\text{peak}-1}$), and an exponential decay for several BX following the peak.  Noise pulse tend to be spread over many BX or localised in one BX. We use the ratios $R_1$ = $E_{\text{peak}+1} / E_{\text{peak}}$ and $R_2$ =  $E_{\text{peak}+2}/E_{\text{peak}+1}$ to characterise the exponential decay, requiring $R_1 > 0.15$ and $0.1 < R_2 < 0.8$. We also require the ratio of the peak energy to the total energy to be between 0.3 and 0.7. Finally, we remove events with more than 30\% of the energy of the pulse outside the central four BX. The noise rejection selection was optimised to maximise the signal to background ratio, using simulated signal and the 2010 control sample, after removal of the cosmic ray background.

\begin{figure}[ht]
\begin{center}
 \includegraphics[width=0.8\linewidth]{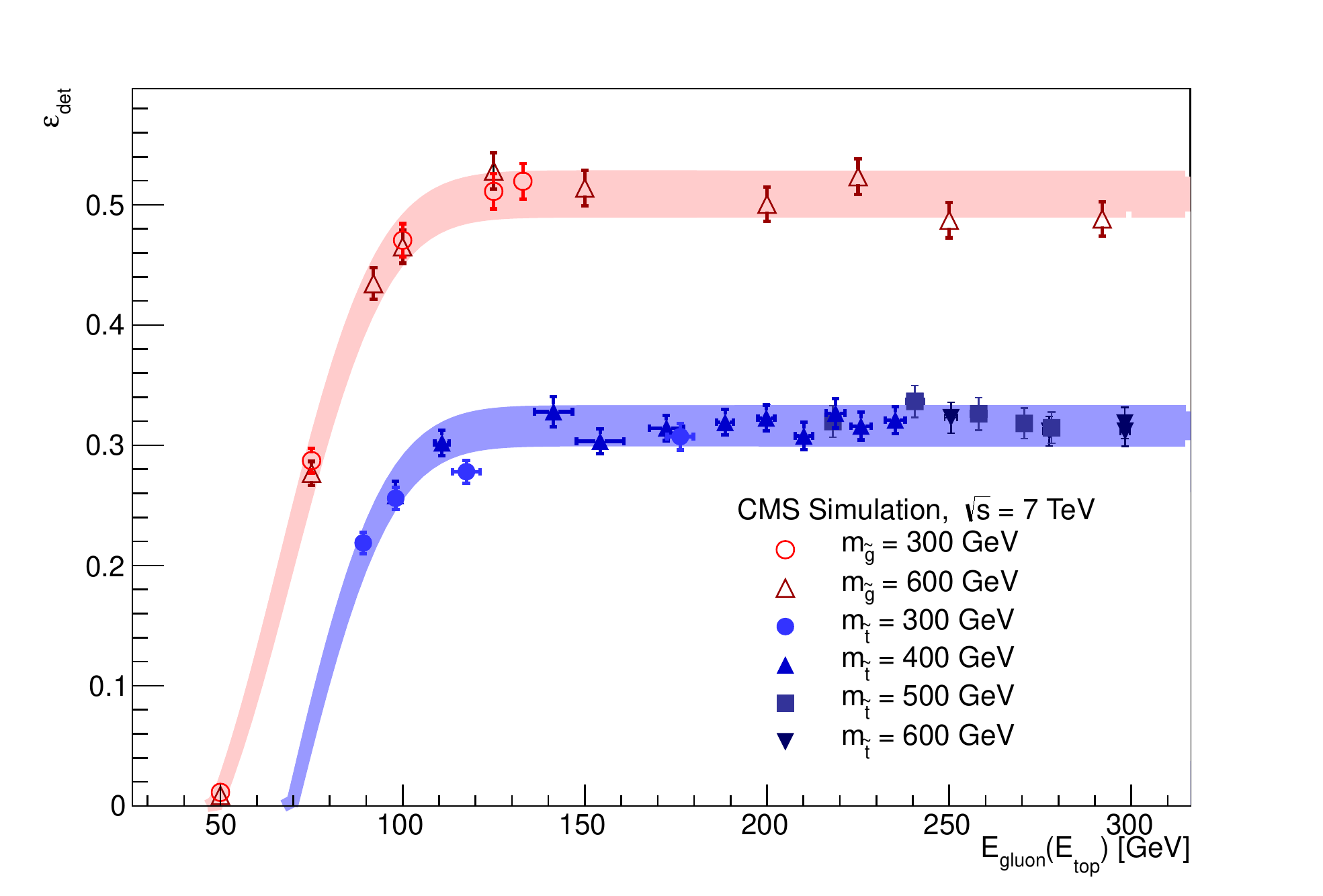}
  \caption{Combined trigger and reconstruction efficiency (detection efficiency) for decays of particles which have stopped in the CMS barrel calorimeters, as a function of the daughter gluon energy, $E_\text{gluon}$, for gluino decays, and as a function of the daughter top energy, $E_\text{top}$, for stop decays.  Each point corresponds to fixed
$m_{\PSg}$ or $m_{\widetilde{\rm t}}$ and $M_{\widetilde{\chi}^0_1}$.  Gluino and stop masses are shown on the plot, while the neutralino masses range from 100--548\GeV for gluino and from 50--300\GeV for stop.
  \label{fig:recoEffScan}}
\end{center}
\end{figure}

After all selection criteria have been applied, the remaining event rate measured in the 2010 control sample is ($5.6 \pm 2.5) \times 10^{-6}$\unit{Hz}. The efficiency for detection of a gluino decay signal (with $m_{\PSg}=600$\GeV and $M_{\widetilde{\chi}^0_1}=490$\GeV), estimated from the simulation as described above, is $(51 \pm 4)$\% of all gluinos stopping in HCAL or ECAL barrel calorimeters. Since we consider a two body decay mode of the gluino, with fixed energy of the daughter gluon, a scan of gluino and neutralino masses allows us to calculate the detection efficiency as a function of $E_\text{gluon}$, as shown in Fig.~\ref{fig:recoEffScan}.  The efficiency is roughly constant provided the gluon energy is above a minimum value, $E^\text{min}_\text{gluon}=100$\GeV. For stop decays, the detection efficiency is lower than the gluino, due to undetectable muons and neutrinos in the final state, but is again roughly constant,
at $(31 \pm 2)$\%, provided the daughter top quark energy is greater than $E^\text{min}_\text{top}=125$\GeV.

\section{Backgrounds}
\label{sec:backgrounds}

The residual backgrounds to this search consist of instrumental noise, unidentified cosmic rays, and unidentified beam backgrounds, principally beam halo.  The cosmic ray muon rejection inefficiency is essentially a geometric factor, which we estimate from the cosmic ray Monte Carlo simulation to be $(0.214 \pm 0.023)\%$.  Taking the product of this factor with the number of events passing the full selection criteria, apart from the cosmic ray veto, we estimate the cosmic ray background to the search to be $5.71 \pm 0.62$ events.  The beam-halo rejection inefficiency is estimated to be approximately 0.2\%, using a ``tag and probe'' method, where the tag is a well identified halo track in one endcap, and the probe is a CSC segment in the other.  The halo background is estimated in bins of jet pseudorapidity, from the product of the rejection inefficiency and the number of positively identified halo events that pass the remaining selection criteria.  The resulting total halo background estimate is $1.50 \pm 0.70$ events.  The noise background rate is assumed to be constant over time, and is calculated from the 2010 control sample.  In this sample we observe 5 events, of which $3.56 \pm 0.39$ are expected to be cosmic rays, giving a noise rate of $(1.5 \pm 2.5)\times 10^{-6}$\unit{Hz}.  The noise background to the search is therefore estimated to be $1.4 \pm 2.2$ events.  The total background to the search is taken as the sum of cosmic ray, halo, and noise backgrounds, giving $8.6 \pm 2.4$ events.

\begin{table}[th]
\caption{Estimated number of background events during the search period for each background source.\label{tab:backgrounds}}
\begin{center} 
\begin{tabular}{|c|c|c|c|}
	\hline 
	Cosmic rays			& Beam-halo 		& Noise					& Total \\ \hline
	$5.71 \pm 0.62$ 	& $1.50 \pm 0.70$	& $1.4 \pm 2.2$	& $8.6 \pm 2.4$ \\
\hline 
\end{tabular}
\end{center} 
\end{table}

\section{Systematic uncertainty}
\label{sec:syst}

A model-independent search for stopped particles avoids many systematic uncertainties common in collider physics, such as those due to parton density functions. However, some sources of systematic uncertainty remain. We assign an uncertainty on the background estimate of 28\%, dominated by the statistical uncertainties in the 2010 control sample used to estimate the noise background, and the sample used to estimate beam-halo background. There is a small systematic uncertainty due to the jet energy scale (JES).   For a JES uncertainty of $\pm10$\%, we estimate a 7\% effect on the cross section limit.  The systematic uncertainty due to trigger efficiency is negligible since the data analysed are well above the turn-on region.  Similarly, the systematic uncertainty due to reconstruction efficiency is negligible since we restrict our search to $E_\text{gluon} >100$\GeV, for gluino, and $E_\text{top} >125$\GeV, for stop.  Finally, there is a 2.2\% uncertainty in the integrated luminosity~\cite{CMS-PAS-SMP-12-008}.

Setting limits on a particular model (e.g. gluinos in split supersymmetry) introduces more substantial systematic uncertainties, since the signal yield is sensitive to the stopping probability.  The {\sc Geant4} simulation used to derive the stopping efficiency described above implements models for both electromagnetic (EM) and nuclear interaction (NI) energy-loss mechanisms.   Whereas the EM model is well understood, the R-hadron cloud model used for the NI has never been tested and is based on postulated physics extrapolated from low-energy QCD.  Moreover, there are alternative models~\cite{Buccella:1985cs} in which R-hadrons preferentially become neutral after a NI.  We do not attempt to quantify these uncertainties; instead we present limits for a particular model.

\section{Search results}
\label{sec:results}

After the selection criteria described in the preceding sections are applied, we perform a counting experiment.  The method is unchanged from that described in Ref. \cite{Khachatryan:2010uf}.  We consider particle lifetime hypotheses from 75\unit{ns} to $10^6$ seconds.  For lifetime hypotheses shorter than one LHC orbit (89\mus),  we search within a time window following each collision, of duration $1.3\times \tau$, for optimal sensitivity to the assumed lifetime $\tau$.  For longer lifetime hypotheses, no search window is used, other than the 2~BX veto around each collision applied during event selection, which affects all lifetime hypotheses.  

For each lifetime hypothesis, the background is assumed to be flat in time, and is calculated from the total live time in which we search, after the time window and the 2~BX veto around collisions have been applied.  An effective integrated luminosity, $L_\text{eff}$,  is calculated for each lifetime, using the Monte Carlo simulation described in Section~\ref{sec:sim}.  The effective luminosity decreases for very short and very long lifetimes, reflecting the fraction of signal that occurs during the search period.  The efficiency for very short lifetimes is reduced by the $\pm2$~BX veto around collisions, and the efficiency for very long lifetimes falls as more signal would appear after the search period is over.

These results of the counting experiment are presented in Table~\ref{tab:results}. In the search sample, we do not observe a significant excess above expected background, $N_\text{exp}$ for any lifetime hypothesis.  It should be noted that the search interval for a given lifetime is either wholly contained within, or equal to, the interval used for any greater lifetime.  Hence the events observed for a small lifetime hypothesis are also observed for longer lifetime hypotheses.  A signal would appear as an increased number of counts across all lifetime search intervals.  For lifetimes shorter than one LHC orbit, the amount of signal in each lifetime bin would allow a measurement of the particle lifetime.  For longer lifetimes, a lifetime measurement could be achieved using dedicated runs after the LHC beams are dumped at the end of each fill.

 \begin{table}[th]
 \topcaption{Results of counting experiments for a range of stopped particle lifetimes, $\tau$.  The effective luminosity, $L_\text{eff}$, for each lifetime is shown, along with the total live time, the expected number of background events, $N_\text{exp}$, and the total number of observed events, $N_\text{obs}$.\label{tab:results}}
 \begin{center} 
 \begin{tabular}{|c|c|c|c|c|}
 	\hline 
 	$\tau$  			& $L_\text{eff}$ (\pbinv) 	& Live time (s)				& $N_\text{exp}$ 			& $N_\text{obs}$ \\ \hline
 	75\unit{ns} 	      		& 19.6					& $2.06 \times 10^4$		&	$ 0.200 \pm 0.056$ 	&	1  \\
 	100\unit{ns} 	            & 57.8					& $6.17 \times 10^4$		&	$ 0.60 \pm 0.17$	&	2  \\
 	1\mus			& 508					& $4.41 \times 10^5$		&	$ 4.3 \pm 1.2$ 		&	7  \\
 	10\mus			& 913					& $8.67 \times 10^5$		&	$ 8.5 \pm 2.4$  	&	12  \\
 	100\mus			& 935					& $8.86 \times 10^5$		&	$ 8.6 \pm 2.4$  	&	12  \\
 	$10^3$\unit{s}			& 866					& $8.86 \times 10^5$		&	$ 8.6 \pm 2.4$ 		&	12  \\	
 	$10^4$\unit{s}			& 636					& $8.86 \times 10^5$		&	$ 8.6 \pm 2.4$ 		&	12  \\
 	$10^5$\unit{s} 			& 332					& $8.86 \times 10^5$		&	$ 8.6 \pm 2.4$  	&	12  \\
 	$10^6$\unit{s} 			& 198					& $8.86 \times 10^5$		&	$ 8.6 \pm 2.4$  	&	12  \\
 \hline 
 \end{tabular}
 \end{center} 
 \end{table}

\begin{figure}[ht]
\begin{center}
  \resizebox{0.9\linewidth}{0.51\linewidth}{\includegraphics{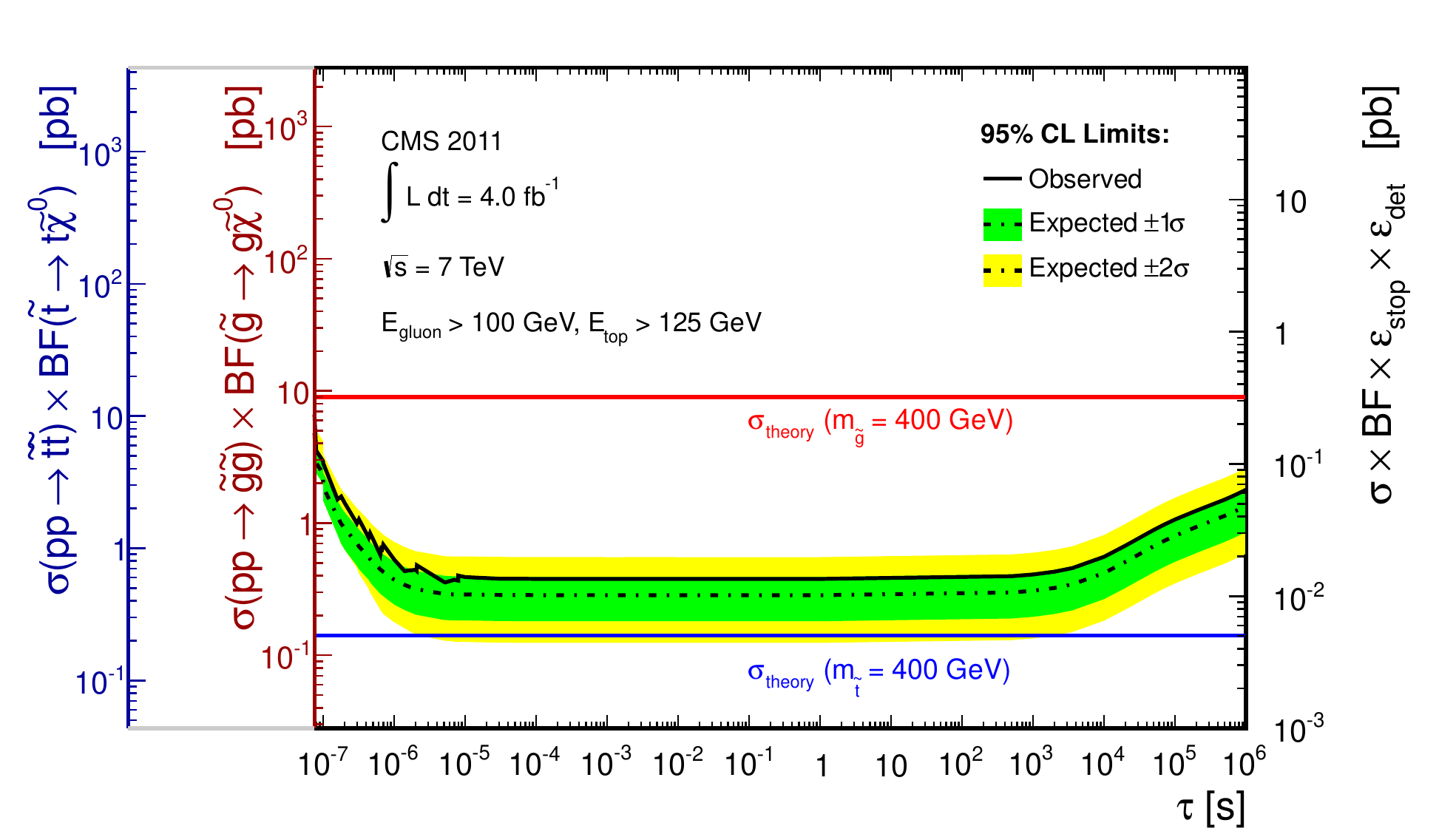}}
  \caption{Expected and observed 95\% CL limits on stop and gluino pair production cross section (left-hand axes), using the cloud model of R-hadron interactions, as a function of particle lifetime.  The theoretical cross sections for 400\GeV gluino and stop production are taken from Ref.~\cite{Beenakker:2011fu}.  Also shown is the model-independent 95\% CL limit on particle production cross-section $\times$ branching fraction $\times$ stopping probability $\times$ detection efficiency (right-hand axis).  The structure observed between $10^{-7}$\unit{s} and $10^{-5}$\unit{s} is due to the number of observed events incrementing when crossing boundaries between lifetime bins.\label{fig:allInOneLifetime}}
\end{center}
\end{figure}

We set 95\%~confidence level (CL) limits over 13 orders of magnitude in stopped particle lifetime, using a CLs~\cite{Junk:1999kv,CMS-NOTE-2011-005} limit calculator implemented in \textsc{ RooStats}~\cite{Moneta:2010pm}.  These limits are presented in Fig.~\ref{fig:allInOneLifetime} as a function of particle lifetime.  The right-hand axis of Fig.~\ref{fig:allInOneLifetime} gives a model-independent limit on particle production cross section $\times$ branching fraction $\times$ stopping probability $\times$ detection efficiency.  We then use the stopping probability and detection efficiency obtained from simulation, to place limits on the particle production cross section, shown on the left-hand axes of Fig.~\ref{fig:allInOneLifetime} for stop and gluino, respectively.  These limits assume visible daughter energies of $E_\text{gluon} > 100$\GeV for gluino, and $E_\text{top} > 125$\GeV for stop, ensuring the detection efficiency is on the plateau shown in Fig~\ref{fig:recoEffScan}.  The sensitivity of the search decreases at short and long lifetimes as the effective luminosity decreases.  The structure observed between $10^{-7}$\unit{s} and $10^{-5}$\unit{s} is due to the number of observed events incrementing across boundaries between lifetime bins.

\begin{figure}[ht]
\begin{center}
  \resizebox{0.8\linewidth}{0.57\linewidth}{\includegraphics{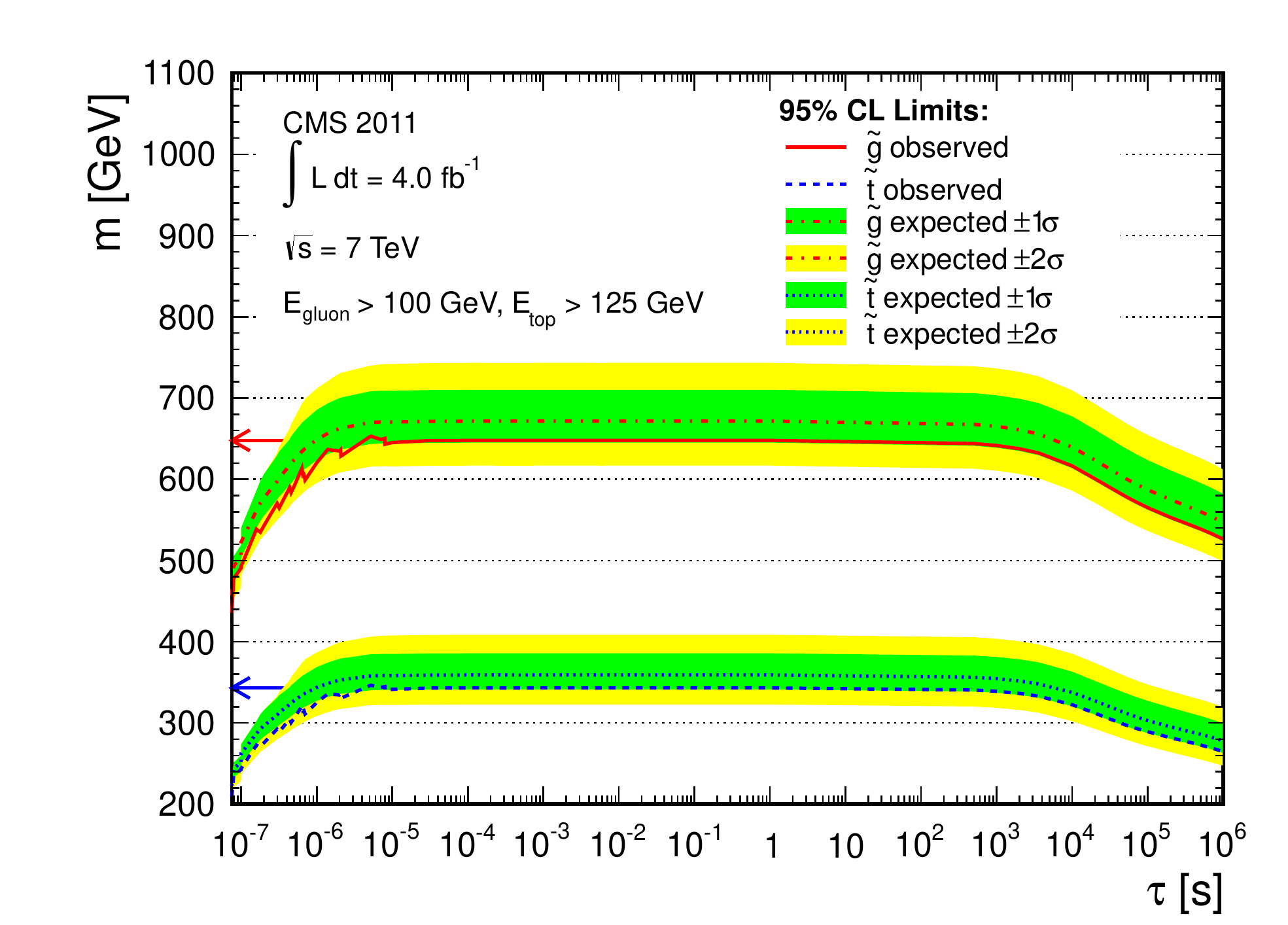}}
  \caption{The 95\% CL limits on gluino and stop mass as a function of particle lifetime, assuming the cloud model of R-hadron interactions and the theoretical production cross sections given in Ref.~\cite{Beenakker:2011fu}.  The structure observed between $10^{-7}$\unit{s} and $10^{-5}$\unit{s} is due to the number of observed events incrementing across boundaries between lifetime bins.  The observed mass limit for the plateau between $10^{-5}$\unit{s} and $10^{3}$\unit{s} is indicated by an arrow on the vertical axis.\label{fig:gluinostopMassLifetime}}
\end{center}
\end{figure}

Figure~\ref{fig:gluinostopMassLifetime} shows the limit on particle mass as a function of lifetime, for gluino and stop, assuming theoretical production cross sections~\cite{Beenakker:2011fu}, as well as BF($\PSg \rightarrow {\rm g}\widetilde{\chi}^0_1$) = 100\% and BF($\widetilde{\rm t} \rightarrow {\rm t}\widetilde{\chi}^0_1$) = 100\%.  For lifetimes between 10\mus to 1000\unit{s}, we exclude gluinos with masses below ${\gluinoMassLimit}$\GeV and stops with masses below ${\stopMassLimit}$\GeV, at 95\% CL.

\section{Excluded region in the \texorpdfstring{$m_{\PSg}$ - $m_{\widetilde{\chi}^{0}}$}{m[gluino] - m[neutralino]} plane}
\label{sec:exclusion}

\begin{figure}[htp]
\begin{center}
  \includegraphics[width=0.8\linewidth]{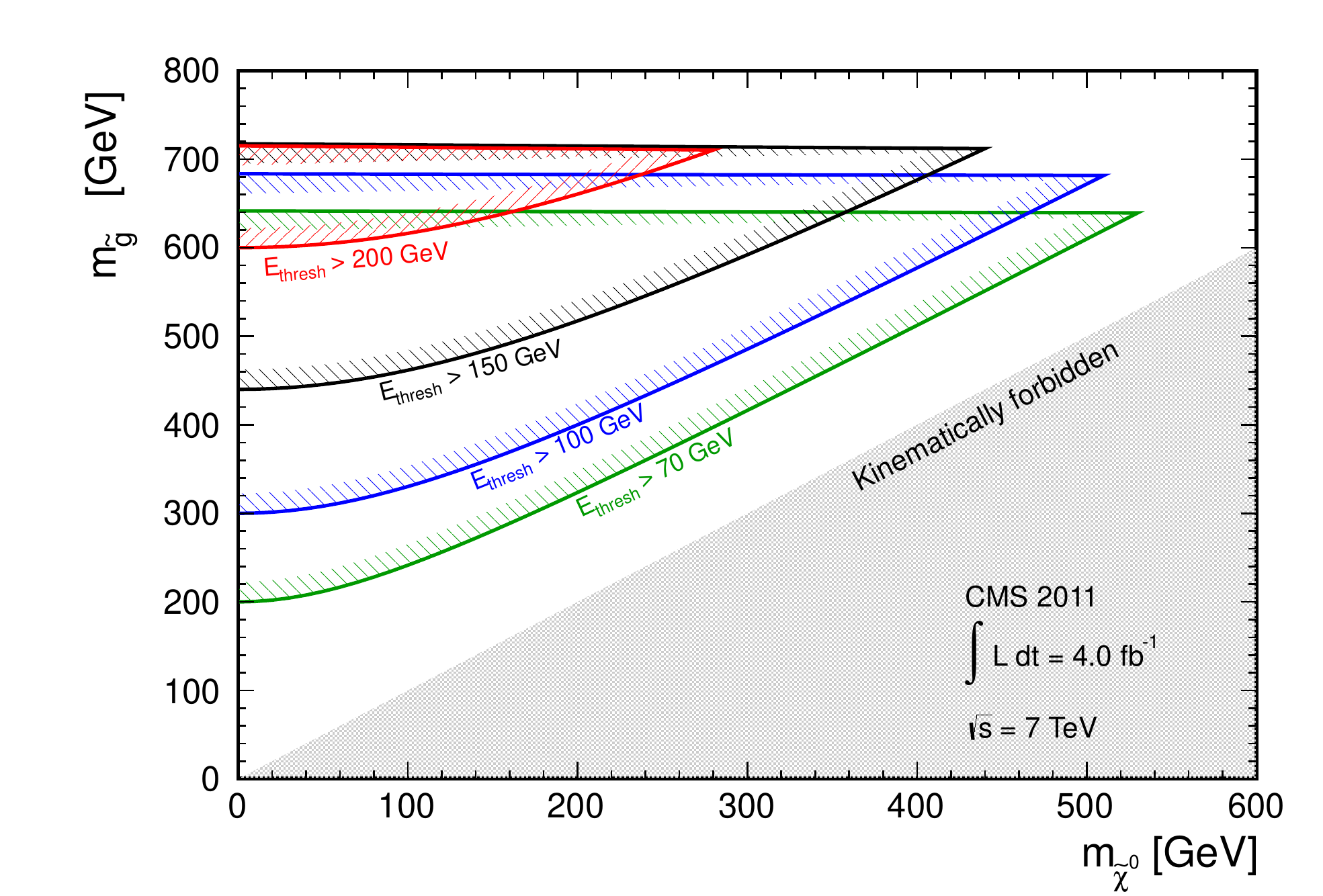}
  \caption{The region of $m_{\PSg}$ - $m_{\widetilde{\chi}^{0}}$ excluded by the analysis, valid for gluino lifetimes $10^{-5}$\unit{s} $< \tau_\text{gluino} <$ $10^{3}$~s, using several jet energy thresholds $E_\text{thresh}$.\label{fig:excludedRegion}}
\end{center}
\end{figure}

For lifetimes in the range 10\mus to 1000\unit{s}, we interpret the results of the analysis as an excluded region in the $m_{\PSg}$ - $m_{\widetilde{\chi}^{0}}$ plane, assuming BF($\PSg \rightarrow {\rm g}\widetilde{\chi}^0_1$) = 100\%.  These results are presented in Fig.~\ref{fig:excludedRegion}.  The excluded region is bounded by two contours, one at constant $m_{\PSg}$ and one at constant $E_\text{gluon}$.  The latter is described by $m_{\PSg}=E_\text{gluon}^\text{min}+\sqrt{{E_\text{gluon}^\text{min}}^{2}+m_{\widetilde{\chi}^{0}}^{2}}$, where
$E_\text{gluon}^\text{min}$ is the minimum gluon energy for which the result is valid, obtained from the start of the plateau in reconstruction efficiency shown in Fig.~\ref{fig:recoEffScan}.

Since the signal efficiency is essentially flat above $E_\text{gluon}^\text{min}$, and the background falls steeply with energy, we obtain stronger limits on the gluino production cross section and hence on $m_{\PSg}$, by repeating the analysis with increased jet energy thresholds ($E_\text{thresh}$) of 100, 150 and 200\GeV.  For each threshold, the background is estimated as described above, and limits are placed on the gluino production cross section and the gluino mass, valid within the .  The results are given in Table~\ref{tab:resultsHighE}, along with the value of $E_\text{gluon}^\text{min}$ for each threshold.  The excluded region of $m_{\PSg}$-$m_{\widetilde{\chi}^{0}}$ is shown separately for each jet energy threshold in Fig.~\ref{fig:excludedRegion}.

\begin{table}[thp]
\setlength{\belowcaptionskip}{6pt}
\topcaption{Results of the analysis using a range of jet energy thresholds.  The expected background, $N_\text{exp}$, and the number of observed events, $N_\text{obs}$, are shown.  The resulting lower limit on the gluino mass ($m_{\PSg}^\text{min}$) assumes the minimum gluon energy ($E_\text{gluon}^\text{min}$) given in the table, gluino lifetime in the range 10~$\mu$s and 1000~s, and BF($\PSg \rightarrow {\rm g}\widetilde{\chi}^0_1$) = 100\%.\label{tab:resultsHighE}}
\begin{center} 
\begin{tabular}{|c|c|c|c|c|c|}
	\hline 
	Threshold (GeV)	& $E_\text{gluon}^\text{min}$ (GeV) & $N_\text{exp}$ & $N_\text{obs}$ & $m_{\PSg}^\text{min}$ (GeV) \\ \hline
	70				& 100			& $8.6 \pm 2.4$		& 12		& 640				\\
	100				& 150			& $3.4 \pm 1.1$		& 5			& 680				\\
	150				& 220			& $1.5 \pm 1.0$		& 2			& 720				\\
	200				& 300			& $0.83 \pm 0.99$	& 2			& 720				\\
\hline 
\end{tabular}
\end{center} 
\end{table}

\section{Summary}

New results have been presented on long-lived particles which have stopped in the CMS detector, after being produced in 7\TeV pp collisions from the CERN Large Hadron Collider.  A search was performed for the decay of such particles, during gaps between LHC beam crossings, using a dedicated calorimeter trigger.  Using a data set in which CMS recorded an integrated luminosity of 4.0\fbinv, and a total search interval of \livetime~hours, a total of 12 events were observed, against a mean background prediction of $8.6 \pm 2.4$ events.  Limits are set at 95\%~CL on long-lived particle pair production, over 13 orders of magnitude of lifetime.  For visible energy $E_\text{gluon} >100$\GeV, assuming BF($\PSg \rightarrow {\rm g}\widetilde{\chi}^0_1$) = 100\%, a gluino with lifetimes ranging from 10\mus to 1000\unit{s} and $m_{\PSg} < {\gluinoMassLimit}$\GeV is excluded.  Under similar assumptions, $E_\text{top} > 125$\GeV, and BF($\widetilde{\rm t} \rightarrow {\rm t}\widetilde{\chi}^0_1$) = 100\%, a stop with lifetimes ranging from 10\mus to 1000\unit{s} and $m_{\widetilde{\rm t}} < {\stopMassLimit}$\GeV is excluded.  By repeating the analysis with increased jet energy thresholds, lower limits on the gluino mass are set up to 720\GeV, valid for $E_\text{gluon} >150$\GeV and lifetimes in the range 10\mus to 1000\unit{s}.  These results considerably extend constraints obtained from previous stopped particle searches~\cite{Abazov:2007ht,Khachatryan:2010uf,Aad:2012zn} and are consistent with the complementary exclusions provided by the direct searches~\cite{Khachatryan:2011ts,Aad:2011hz,Chatrchyan:2012sp}.

\section*{Acknowledgements}
We congratulate our colleagues in the CERN accelerator departments for the excellent performance of the LHC machine. We thank the technical and administrative staff at CERN and other CMS institutes, and acknowledge support from: FMSR (Austria); FNRS and FWO (Belgium); CNPq, CAPES, FAPERJ, and FAPESP (Brazil); MES (Bulgaria); CERN; CAS, MoST, and NSFC (China); COLCIENCIAS (Colombia); MSES (Croatia); RPF (Cyprus); MoER, SF0690030s09 and ERDF (Estonia); Academy of Finland, MEC, and HIP (Finland); CEA and CNRS/IN2P3 (France); BMBF, DFG, and HGF (Germany); GSRT (Greece); OTKA and NKTH (Hungary); DAE and DST (India); IPM (Iran); SFI (Ireland); INFN (Italy); NRF and WCU (Korea); LAS (Lithuania); CINVESTAV, CONACYT, SEP, and UASLP-FAI (Mexico); MSI (New Zealand); PAEC (Pakistan); MSHE and NSC (Poland); FCT (Portugal); JINR (Armenia, Belarus, Georgia, Ukraine, Uzbekistan); MON, RosAtom, RAS and RFBR (Russia); MSTD (Serbia); MICINN and CPAN (Spain); Swiss Funding Agencies (Switzerland); NSC (Taipei); TUBITAK and TAEK (Turkey); STFC (United Kingdom); DOE and NSF (USA).  Individuals have received support from the Marie-Curie programme and the European Research Council (European Union); the Leventis Foundation; the A. P. Sloan Foundation; the Alexander von Humboldt Foundation; the Belgian Federal Science Policy Office; the Fonds pour la Formation \`a la Recherche dans l'Industrie et dans l'Agriculture (FRIA-Belgium); the Agentschap voor Innovatie door Wetenschap en Technologie (IWT-Belgium); the Council of Science and Industrial Research, India; and the HOMING PLUS programme of Foundation for Polish Science, cofinanced from European Union, Regional Development Fund.

\clearpage

\bibliography{auto_generated}

\cleardoublepage \appendix\section{The CMS Collaboration \label{app:collab}}\begin{sloppypar}\hyphenpenalty=5000\widowpenalty=500\clubpenalty=5000\input{EXO-11-020-authorlist.tex}\end{sloppypar}
\end{document}

%% file: EXO-11-020-authorlist.tex
\textbf{Yerevan Physics Institute,  Yerevan,  Armenia}\\*[0pt]
S.~Chatrchyan, V.~Khachatryan, A.M.~Sirunyan, A.~Tumasyan
\vskip\cmsinstskip
\textbf{Institut f\"{u}r Hochenergiephysik der OeAW,  Wien,  Austria}\\*[0pt]
W.~Adam, T.~Bergauer, M.~Dragicevic, J.~Er\"{o}, C.~Fabjan\cmsAuthorMark{1}, M.~Friedl, R.~Fr\"{u}hwirth\cmsAuthorMark{1}, V.M.~Ghete, J.~Hammer, N.~H\"{o}rmann, J.~Hrubec, M.~Jeitler\cmsAuthorMark{1}, W.~Kiesenhofer, V.~Kn\"{u}nz, M.~Krammer\cmsAuthorMark{1}, D.~Liko, I.~Mikulec, M.~Pernicka$^{\textrm{\dag}}$, B.~Rahbaran, C.~Rohringer, H.~Rohringer, R.~Sch\"{o}fbeck, J.~Strauss, A.~Taurok, P.~Wagner, W.~Waltenberger, G.~Walzel, E.~Widl, C.-E.~Wulz\cmsAuthorMark{1}
% ...
\vskip\cmsinstskip
\dag:~Deceased\\
1:~~Also at Vienna University of Technology, Vienna, Austria\\
% ...
